\documentclass[pra,aps,groupedaddress,superscriptaddress,notitlepage, twocolumn]{revtex4-2}

\usepackage[utf8x]{inputenc}
\usepackage{color}
\usepackage{bbm}

\usepackage{nicefrac}
\usepackage{amsfonts,amsmath,amssymb,stmaryrd}
\usepackage{braket}

 \usepackage[autostyle=true]{csquotes}

\usepackage{tabularx}
\usepackage{multirow} 
\usepackage{hhline}
\usepackage{graphicx}
\usepackage{subfigure}  
\usepackage{bbm} 
\usepackage{hyperref}
\usepackage[pdftex]{epsfig}
\usepackage{mathrsfs}
\usepackage{verbatim}
\usepackage{centernot}
\usepackage{array}
\usepackage{cancel}
\usepackage{ifthen}
\usepackage{bm}	
\usepackage{siunitx}
\usepackage{float} 
\usepackage[super]{nth} 
\usepackage{todonotes}
\usepackage{color,soul}
\InputIfFileExists{gitinfo-latex.inc}{}{}

\usepackage{listings}
\usepackage{color}

\usepackage[acronym,nomain]{glossaries}

\usepackage{xcolor}
\newcommand{\el}[1]{\textcolor{black}{#1}}

\begin{document}
	
\title{Power of a quasi-spin quantum Otto engine at negative effective temperature}
\author{Jens Nettersheim}
\affiliation{Department of Physics and Research Center OPTIMAS, Technische Universit\"at Kaiserslautern, Germany}

\author{Sabrina Burgardt}
\affiliation{Department of Physics and Research Center OPTIMAS, Technische Universit\"at Kaiserslautern, Germany}

\author{Quentin Bouton}
\affiliation{Department of Physics and Research Center OPTIMAS, Technische Universit\"at Kaiserslautern, Germany}
\affiliation{Laboratoire de Physique des Lasers, CNRS, UMR 7538, Université Sorbonne Paris Nord, F-93430 Villetaneuse, France}

\author{Daniel Adam}
\affiliation{Department of Physics and Research Center OPTIMAS, Technische Universit\"at Kaiserslautern, Germany}

\author{Eric Lutz}
\affiliation{Institute for Theoretical Physics I, University of Stuttgart, D-70550 Stuttgart, Germany}

\author{Artur Widera}
\email{email: widera@physik.uni-kl.de}
\affiliation{Department of Physics and Research Center OPTIMAS, Technische Universit\"at Kaiserslautern, Germany}
\affiliation{Graduate School Materials Science in Mainz, Gottlieb-Daimler-Strasse 47, 67663 Kaiserslautern, Germany}	

\date{\today}

\begin{abstract}
Heat engines usually operate by exchanging heat  with thermal baths at different (positive) temperatures.  Nonthermal baths may, however, lead to a significant performance boost.   We here experimentally analyze the power output of a  single-atom quantum Otto engine realized in the quasi-spin states of individual Cesium atoms interacting with an atomic Rubidium bath.  From measured time-resolved populations of the quasi-spin state,  we determine  the dynamics during the  cycle of both the effective spin temperature and of the quantum fluctuations of the engine, which we quantify with the help of the Shannon entropy.  We find that power is enhanced in the negative temperature regime, and that it reaches its maximum value at half the maximum entropy.  
Quantitatively, operating our engine at negative effective temperatures increases the power by up to $30\%$ compared to operation at positive temperatures, including even the case of infinite temperature. \el{At the same time, entering the negative temperature regime allows for reducing the entropy to values close to zero, offering highly stable operation at high power output.} \el{We furthermore numerically investigate the influence of the size of the Hilbert space  on the performance of the quantum engine by varying the number of levels of the working medium. \el{Our work thereby paves the way to fluctuation control in the operation of high-power and efficient single-atom quantum engines.}}
\end{abstract}

\maketitle

\section{Introduction}

The capability of heat engines to produce mechanical work out of  thermal energy has played a pivotal role in our society. They have  been extensively utilized to generate motion, ranging from vehicles and ships to trains and airplanes \cite{cen01}. Two central figures of merit of heat engines are efficiency, defined as the ratio of work output and heat input, and power that characterizes the work output rate. Large power output is essential  from a practical point of view. Thermal machines with vanishing power  have indeed limited application for transportation purposes, even if they have  very high efficiency \cite{cen01}. 
Macroscopic engines are commonly coupled to two (cold and hot) heat reservoirs that are used to modify the temperature of the working medium. Such large heat baths usually lead to thermal states with positive temperature.

 Technical development in the past decades has lead to a rapid miniaturization of thermal machines. The realization of both classical \cite{hug02,sta11,bli12,mar15,ros16,lin19,hor20} and quantum \cite{zou17,kla19,pet19,bou20} heat engines has been reported. These microscopic engines are frequently coupled to engineered reservoirs that can be easily controlled. As a result, they offer the possibility to prepare unconventional baths, such as coherent \cite{scu03,scu11}, squeezed \cite{aba14a,kla17} or negative-temperature \cite{ass19} reservoirs. Remarkably, nonthermal baths have been predicted to enhance the performance of (classical and quantum) heat engines \cite{scu03,scu11,aba14a,kla17,ass19,aba14,ali15}. They can thus be considered as a useful thermodynamic resource. Recently, an efficiency boost has been experimentally demonstrated for a classical nanomechanical engine coupled to a squeezed reservoir \cite{kla17} and a quantum two-level engine interacting with a bath at negative effective temperature \cite{ass19}.
 
The study of  negative temperatures has a long experimental \cite{pur51,oja97,med11,bra13} and theoretical  \cite{ram56,lie66,mos05,rap10,dun14,str18,ham18,bal21} tradition, most prominently in nuclear magnetic resonance and atomic physics. They appear in systems with upper bounded energy spectra and are associated with inverted  states, known, for instance, from population inversion in laser physics \cite{scu97}.  They correspond to  occupation probabilities that  increase with the energy of the
state --- and not decrease with it, as is the case for thermal states with positive temperatures. Consequently, a system with a negative temperature is more energetic than one with (positive)  infinite temperature \cite{kit80}. States with apparent negative temperatures may be interpreted as unstable nonequilibrium states that are consistent with the standard   second law {of thermodynamics} \cite{str18}.
 
 In this paper, we experimentally investigate the occurrence of effective negative temperatures, and their enhancing effect on the power output,  of a quantum Otto cycle  consisting of a large quasi-spin (with $N=7$ levels)  of individual Cesium (Cs) atoms collisionally coupled to an ultracold Rubidium (Rb) gas, which plays the role of a quantum heat bath \cite{bou20}. Expansion and compression  are implemented by modulating an external magnetic field, that changes the energy-level spacing of the engine via the Zeeman effect, and thus performs work \cite{kos17}. On the other hand, heat exchange between system and {cold (hot)} \el{spin} bath occurs via inelastic endoenergetic {(exoenergetic)}  spin-exchange collisions \cite{Schmidt2019}. We employ quantum control of the coherent spin-exchange process \cite{sik18} to control the direction of heat transfer  at the level of individual quanta, and hence realize heating and cooling \cite{kos17}. We additionally use single-atom and time-resolved measurements \cite{Bouton2020} of the quasi-spin distribution of a Cs atom to monitor the population dynamics of the engine along the cycle. As a result, we are able to follow the time evolution of both the effective spin temperature and of the power output, and analyze their relationship. This allows us to gain unique insight into the inner workings of a nanoscopic {single-atom} quantum heat engine.

 The structure of the paper is as follows. We start in Sec.~II by describing the experimental setup and the implementation of the quantum Otto cycle. In Sec.~III, we analyze the population dynamics of the  engine and determine the effective spin temperature during the cycle. We further investigate the quantum fluctuations of the system by evaluating the  Shannon entropy of the spin distribution of the engine. In Sec.~IV, we examine the time evolution of the power output and observe a power boost associated with the effective negative temperature regime. Finally, in Sec.~V, we compare the performance of our seven-level engine with that of \el{a $N$-level} machine.
 
\section{Experimental system}

\begin{figure}[t]
\centering
\includegraphics[width=8.5cm]{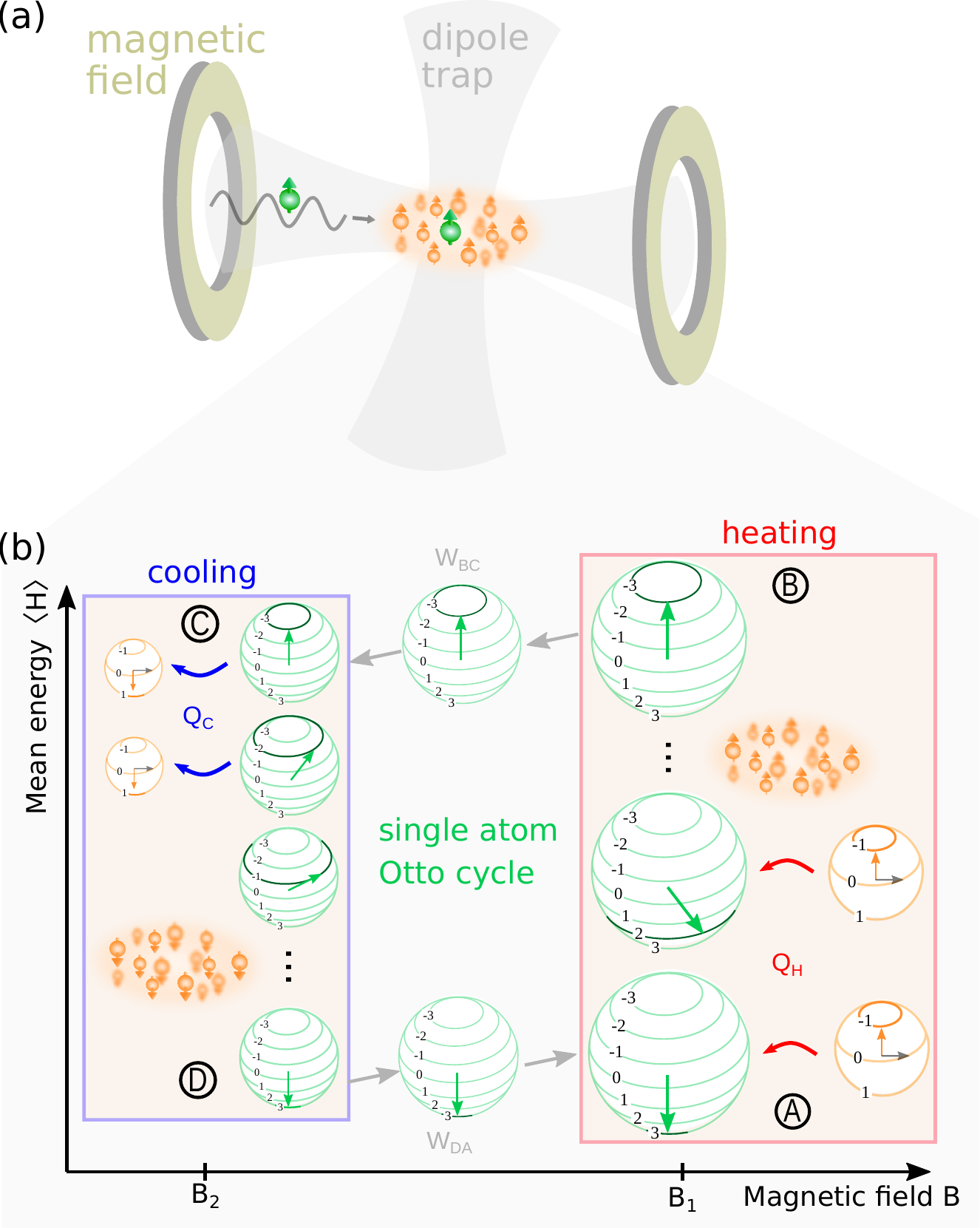}
\caption{{Quasi-spin quantum Otto engine with the experimental setup. (a) Transport of a single Cs atom (green) into the ultracold Rb bath (orange) via an optical conveyor belt lattice (sinuous line). Rb is trapped in an optical crossed dipole trap illustrated in grey. Magnetic field coils (yellow) control the magnetic environment. (b) The cycle of our quantum Otto heat engine is made of four strokes: heating via spin-exchange collisions with the Rb bath (A $\rightarrow$ B) which provides heat $\langle Q_\text{H}\rangle$,  compression by a magnetic field ramp (B $\rightarrow$ C),  decreasing the energy level splitting of the Cs atoms, cooling via spin-exchange collisions with the Rb bath (C $\rightarrow$ D) which extract heat $\langle Q_\text{C}\rangle$, and compression by a magnetic field ramp (D $\rightarrow$ A) that increases the energy level splitting back to its initial value. In the figure, $m_F$ states are shown as level schemes of green (Cs) and orange (Rb) spheres. Gray arrows in Rb spheres indicate the $m_F=0$ state, which acts as the exhaust gas of our quantum heat engine.}}
\label{fig:Complete_cycle_Blochspheres_v1}
\end{figure}

\el{In our experiment, we} immerse few Cs atoms into a bath of $10^4$  Rb atoms at a kinetic temperature of approximately $T=1\,\mu$K --- both species are confined in a common optical dipole trap \cite{bou20}, as {illustrated in Fig.~\ref{fig:Complete_cycle_Blochspheres_v1}(a)}.  
The engine is realized in the seven hyperfine ground states of Cs $\ket{F_\text{Cs}=3,m_{F,\text{Cs}}}$ with the magnetic substates $m_{F,\text{Cs}}\in [+3, +2, \ldots, -3]$ (the lowest-energy state $m_{F,\text{Cs}} = 3$ marks the zero-point).
The bath of Rb atoms is prepared in the state $\ket{F_\text{Rb}=1,m_{F,\text{Rb}}}$ with $m_{F,\text{Rb}}= \pm 1$. 
Spin-exchange collisions realize the heat transfer between engine and bath, where in every collision the $m_F$ state {of the colliding Rb and Cs atom} is changed:
A transition of the engine's state by $\Delta m_{F,\text{Cs}}=-1$ ($+1$) corresponds to a directed energy transport where energy is absorbed (emitted) by the Cs atom. The strong atom number imbalance of $N_\text{Rb}/N_\text{Cs}>1000$ ensures that a bath atom collides only once with a Cs atom and is therefore transferred to $m_{F,\text{Rb}}= 0$; the bath can thus be considered as Markovian. \el{Importantly, the engine is driven purely by the "spin fuel" of the Rb bath, while the kinetic temperature remains constant at all times. The effective spin temperature of this quantum bath is close to $T=\pm 0\,$K, where the sign depends on the spin orientation. The bath therefore features a temperature range not accessible by usual thermal baths.}
During the engine cycle,  the magnetic field $B$ changes between $B_1 = 346.5 \pm 0.2\,$mG and $B_2 = 31.6 \pm 0.1\,$mG.
The Cs energy splitting at these magnetic fields is given by $E_n^\text{Cs}= n \lambda B$ with $n = 3-m_{F,\text{Cs}}$ ($n \in 0, 1, ..., 6$ corresponding to $m_F \in 3, 2, ..., -3$) and $\lambda = |g_F^\text{Cs}| \mu_B$ where $\mu_B$ is the Bohr magneton and $g_F^\text{Cs}= -1/4$ the Cs Landé factor. 
The amount of energy provided by a Rb atom during a spin-exchange collision  is $\Delta E^\text{Rb} = \mp \kappa B$,  with $\kappa= |g_F^\text{Rb}| \mu_\text{B}$, where $g_F^\text{Rb}=-1/2$ is the Rb Landé factor \cite{Schmidt2019}. 
This value is twice the amount in the Cs spin system due to the twice larger Rb Landé factor. Spin-exchange collisions are thus inelastic. While each single collision is coherent and thus
 amenable to quantum state engineering, the coupling  to the large number of bath modes in inelastic
 collisions destroys the coherence between the engine's
 quasi-spin levels.

The quantum Otto cycle consists of four different branches  \cite{kos17}: one compression and one expansion step, during which work is performed, as well as a heating and a cooling stage,  during which heat is exchanged (Fig.~\ref{fig:Complete_cycle_Blochspheres_v1}(b)).
Heat, $\langle Q\rangle =\sum_n E_n^\text{Cs} \Delta p_n$, is transferred between engine and bath, via inelastic spin-exchange collisions, during  heating (A $\rightarrow$ B) and  cooling (C $\rightarrow$ D)  strokes. 
\el{Compression} (D $\rightarrow$ A) and  expansion (B $\rightarrow$ C) strokes, realized as adiabatic magnetic field changes, perform work, $\langle W \rangle= \sum_n p_n \Delta E_n^\text{Cs}$, where $p_n$ is the Cs population in the $n$-th $m_F$ state.  The cycle time is defined as $\tau_\text{cycle} = \tau_\text{H}+\tau_\text{C} + 2 \tau$, where $\tau_\text{H,C}$ are the durations of the heating and cooling phases, and $\tau$ that of \el{both} the adiabatic expansion and compression steps. In order to evaluate the above quantities,
we determine the  magnetic fields $B_1$ and $B_2$  with the help of Rb microwave spectroscopy \cite{bou20}. We additionally detect the Zeeman populations $p_n^i$ of individual Cs atoms at arbitrary times by position resolved fluorescence measurements combined with Zeeman-state selective operations \cite{Bouton2020}. 
From a series of such measurements, we can, atom by atom, construct the quasi-spin populations at any time during the heat engine cycle (Fig.~2, inset). 

Our system possesses several unique features that  distinguish it from other microscopic machines \cite{hug02,sta11,bli12,mar15,ros16,lin19,hor20,zou17,kla19,pet19}. First, the engine spans  seven quantum states exhibiting a natural stop of heat exchange if the highest or lowest energy state is populated. This is in marked contrast to unbound energy spectra such as harmonic oscillator systems.  This property allow us to control  the quantum fluctuations in the system,  which are associated
 with random transitions between its discrete levels. Second, the quantum bath is realized by an ultracold atomic gas, where the spin polarization rather than the kinetic temperature stores the fuel of the engine. Controlling coherent collisional events, we can  direct the heat exchange to either  increase or decrease the engine's spin energy in each spin-exchange collision.  This aspect permits us to  inverse the engine's spin population and achieve  negative spin temperatures. 
Third, we can resolve the quantum-spin population time evolution  at the level of individual quanta,  yielding access to the mean but also to the fluctuations of the heat energy of the engine. This feature enables us to follow the dynamics of the heat engine throughout its cycle.

\begin{figure}[t]
	\centering
	\includegraphics[width=8.5cm]{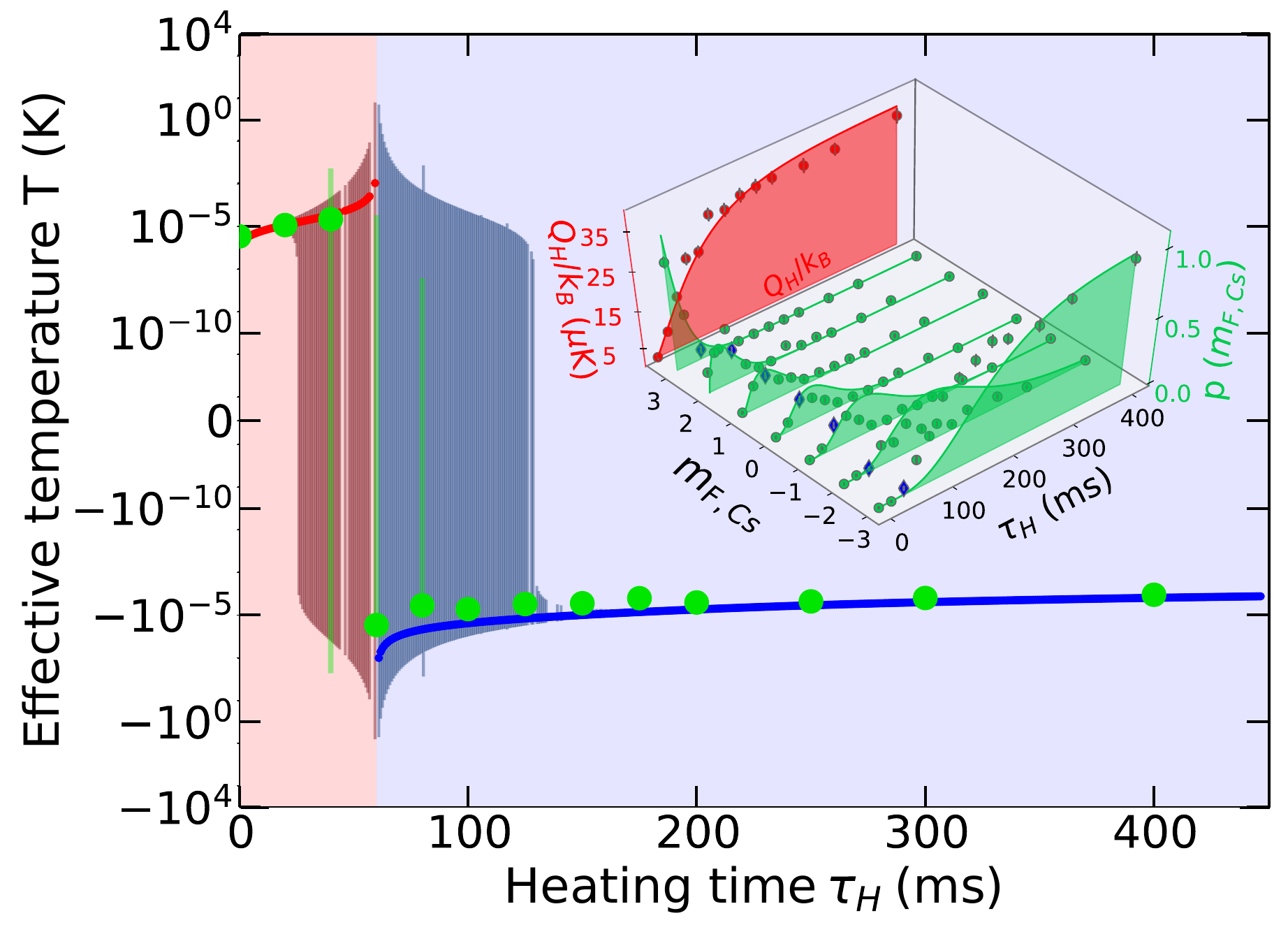}
	\caption{{Evolution of the effective spin temperature. {By fitting the Cs population with $P_n = a  P_{+} + (1-a) P_{-}$ and Boltzmann distributions \el{$P_\pm = \exp(-E_n^\text{Cs}/k_B T_\pm)/Z$},  a positive (red) and negative (blue) temperature is \el{extracted}. The presented positive or negative temperature has a fit contribution of $98\%$.  Large error bars indicate the transition area from positive to effective negative temperature. Green dots represent extracted effective temperatures from measurements. Background colors indicate positive (red) and negative (blue) effective temperature \el{regimes}.} The inset shows the time evolution of the spin population and exchanged heat $Q_H$ of the heat engine reported in Ref.~\cite{bou20}.}
	}
	\label{fig:spin_time_evolution}
\end{figure}

\section{Population dynamics and negative spin temperatures}

Combining quantum engineering of the machine's  and bath's spin states, we can control  the energy transfer between the engine and the atomic bath at the level of single quanta. Starting from the lowest Cs energy level {$\ket{m_{F,\text{Cs}}=3}$}, a maximum of six quanta can be stored in the machine {after}  six spin-exchange collisions during heating, after which the system is in the highest Cs energy level {$\ket{m_{F,\text{Cs}}=-3}$}. Conversely,  starting from the highest Cs energy level, a maximum of six quanta can be extracted from the machine {after} six spin-exchange collisions during cooling, after which the system is again in the lowest Cs energy level.

The population dynamics of the engine and the corresponding time-resolved heat transfer is shown in the inset of Fig.~\ref{fig:spin_time_evolution} for a typical heating  phase.  Starting from the lowest-energy, spin-polarized state, spin-exchange collisions significantly broaden the distribution $p(m_{F,\text{Cs}})$ until after approximately three spin-exchange collisions a maximally broad distribution is reached. Further heat transfer leads to a reduction of the distribution's width until the other state of extreme energy is reached. This population dynamics will significantly affect the effective spin temperature and the size of quantum fluctuations in the system --- and, in turn, the thermodynamic performance of the machine.

In order to quantify the effective spin temperature, we fit a sum of two Maxwell-Boltzmann distributions to the data, $P_n = a  P_{+} + (1-a) P_{-}$,  one for positive ($+$) and the other for negative ($-$) temperatures, where 

\begin{align}
	P_\pm= \exp(-E_n^\text{Cs}/k_B T_\pm)/Z,
\end{align}

with $Z$ the partition function and $T_\pm$ the effective positive or negative temperature ($k_\text{B}$ is the Boltzmann constant). The amplitude $a$ is a fit parameter which allows us to identify regimes of positive or negative temperatures or, for large uncertainty of $a$ in a fit, the transition regime (see Appendix A for details). The time evolution of the effective spin temperature  is shown in Fig.~\ref{fig:spin_time_evolution} as a function of the heating time $\tau_\text{H}$. For short interaction times, the data (green dots) is well reproduced by an effective positive temperature, which increases with interaction time (solid, blue and red, lines are simulations of a theoretical model, see Appendix B).  The transition between positive and negative temperatures occurs  at roughly $60$ms, when the spin distribution shows its largest width.  For longer interaction times,  the system is well described by a negative effective temperature, indicating population inversion \cite{scu97}.  The largest amount of energy turnover during heating  is reached for an initial (final) state with very small  effective temperature in the positive (negative) temperature domain.

\begin{figure}
	\centering
	\includegraphics[width=8.3cm]{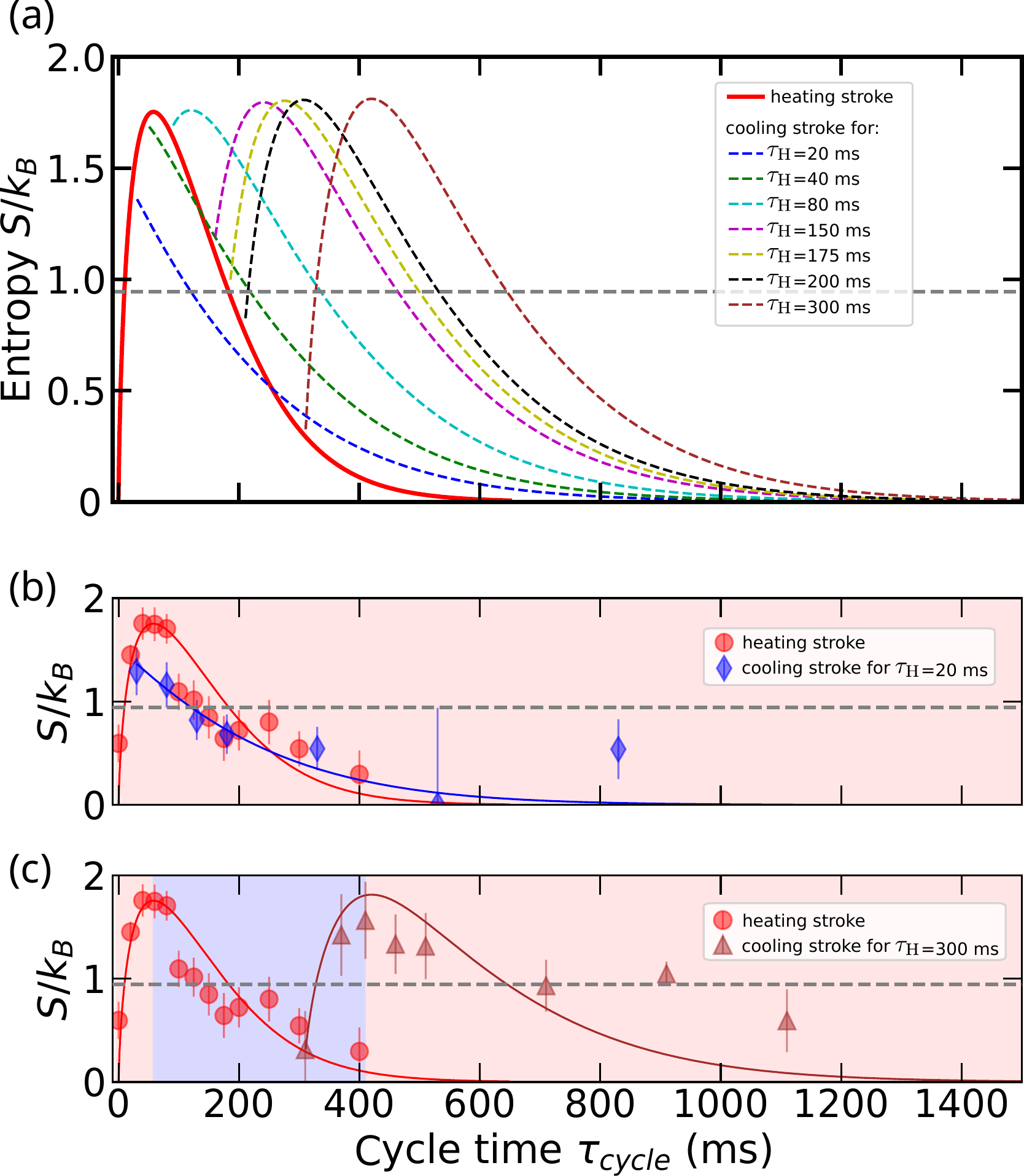}
	\caption{{Evolution of the Shannon entropy. {(a) Shannon entropy, Eq.~(2), as a function of the cycle time $\tau_\text{cycle}$: behavior for the heating stroke A $\rightarrow$ B (red, solid) and for the cooling stroke  C $\rightarrow$ D (other colors, dashed). Times in the legend indicate the heating time $\tau_\text{H}$.
	The initial state $\ket{m_{F,\text{Cs}}=+3}$ is fully polarized (see Fig.~\ref{fig:spin_time_evolution}) and leads to an initial zero entropy. Large heating times populates the system in $\ket{m_{F,\text{Cs}}=-3}$ giving a low entropy again. The maximum is reached for a mostly spread spin distribution, theoretically reachable for a population of $1/7$ in each of the seven states. {Populating highly excited states during heating leads to a second entropy maximum while cooling.}
	Short heating does not lead to the double peak structure. 
	(b) Evolution of the Shannon entropy for a small heating time $\tau_\text{H} = 20$ms. Dots show data of the heating stroke and diamonds for the cooling stroke.  Solid lines are simulations based on our spin-exchange model.
	 Horizontal dashed lines mark the entropy of cycle points B and C when the engine runs at full power.} 
	(c) Evolution of the Shannon entropy for a large heating time $\tau_\text{H} = 300$ms. Dots show data of the heating stroke and triangles for the cooling stroke One can see that for the larger time the system operates in the negative temperature regime between the two entropy peaks, which is not crossed for the shorter one. }}
		\label{fig:entropy_evolution}
\end{figure}

The population dynamics impacts the size of quantum fluctuations in the system. Fluctuations vanish in the lowest and highest  energy states since they are pure, and increase for mixed states. We quantify the evolution of the width of the engine's spin distribution, and thus of the quantum fluctuations, by computing the Shannon entropy of system  during the heat-exchange process. From the measured population distribution of each $m_F$ state at every time, the Shannon entropy is given by \cite{kit80}
\begin{equation}
	S = - k_{\text{B}} \sum_{m_F=+3}^{-3} p_{m_F} \ln{p_{m_F}}.
	\label{eq:entropy}
\end{equation}
Entropies have been extensively used to quantify the size of fluctuations, most notably in entropic uncertainty relations \cite{maa88,col12,col17}. They have been shown to offer better bounds than those based on the variance \cite{maa88,col12,col17}. We here employ Eq.~\eqref{eq:entropy} to quantify (i) the size of quantum fluctuations \cite{maa88,col12,col17} and (ii) the amount of information \cite{kit80} stored in the engine.

The entropy dynamics is shown as a function of the cycle time, $\tau_\text{cycle} = \tau_\text{H}+\tau_\text{C} + 2 \tau$, in Fig.~\ref{fig:entropy_evolution}(a) as a solid red line for the heating stage (A $\rightarrow$ B) (the expansion/compression time $\tau$ is held constant and the heat/cooling time is varied).  It has been calculated from numerical simulations of the heating process  using the experimental parameters (see Appendix B for details {of the spin-exchange model}).  Starting from the lowest-energy, zero-entropy state, the entropy increases as more engine-spin states are populated by spin-exchange collisions. The entropy passes a maximum, which is close to the {theoretical} maximum-entropy value for seven spin states of $S_\mathrm{max}=k_B \ln 7$ at approximately $60\,$ms, until it approaches zero for longer interaction times. Interestingly, the entropy maximum (largest quantum fluctuations/stored information) occurs at the time at which the effective temperature becomes negative (a similar behavior is known for the case of a two-level system \cite{kit80}).
At the end of a full heating step, the system is in the maximum-energy, zero-entropy state, and the subsequent cooling stroke starts with the polarized state $\ket{m_{F,\text{Cs}}=-3}$. However, the entropy evolution during the cooling stage ($\text{C} \rightarrow \text{D}$) (dashed colored lines in Fig.~\ref{fig:entropy_evolution}(a)) depends strongly on whether the maximum entropy point, {located at $\tau_\text{H}=\SI{58}{\milli\second}$}, has been reached  or not: For short heating rates ({$\tau_\text{H} \leq \SI{58}{\milli\second}$}, the entropy decreases monotonically to zero --- in this case, the entropy curve displays a single peak. For long heating rates ({$\tau_\text{H} \geq \SI{58}{\milli\second}$}, the entropy first increases, reaches its maximum value, before decreasing again  --- in this case, the entropy curve exhibits a double-peak structure. 

Both behaviors are observed experimentally (see Figs.~\ref{fig:entropy_evolution}(b)-(c)). The data for full heating are represented by red dots and agree well with the simulated curve (deviations are due to experimental imperfections). For a short heating time  ($\tau_\text{H} =20$ms) (blue diamonds in Fig.~\ref{fig:entropy_evolution}(b)), the negative temperature domain is not reached, and the entropy decreases monotonically during cooling. By contrast, for a long heating time  ($\tau_\text{H} =300$ms) ({brown} triangles in Fig.~\ref{fig:entropy_evolution}(c)), the negative temperature domain is  reached (depicted by blue shaded area), and the entropy curve exhibits two distinct peaks. {After reaching the second entropy peak, most atoms populate low energetic states, and the engine enters the effective positive temperature operation area, highlighted by the red background.}

\begin{figure}[t]
	\centering
	\includegraphics[width=9.3cm]{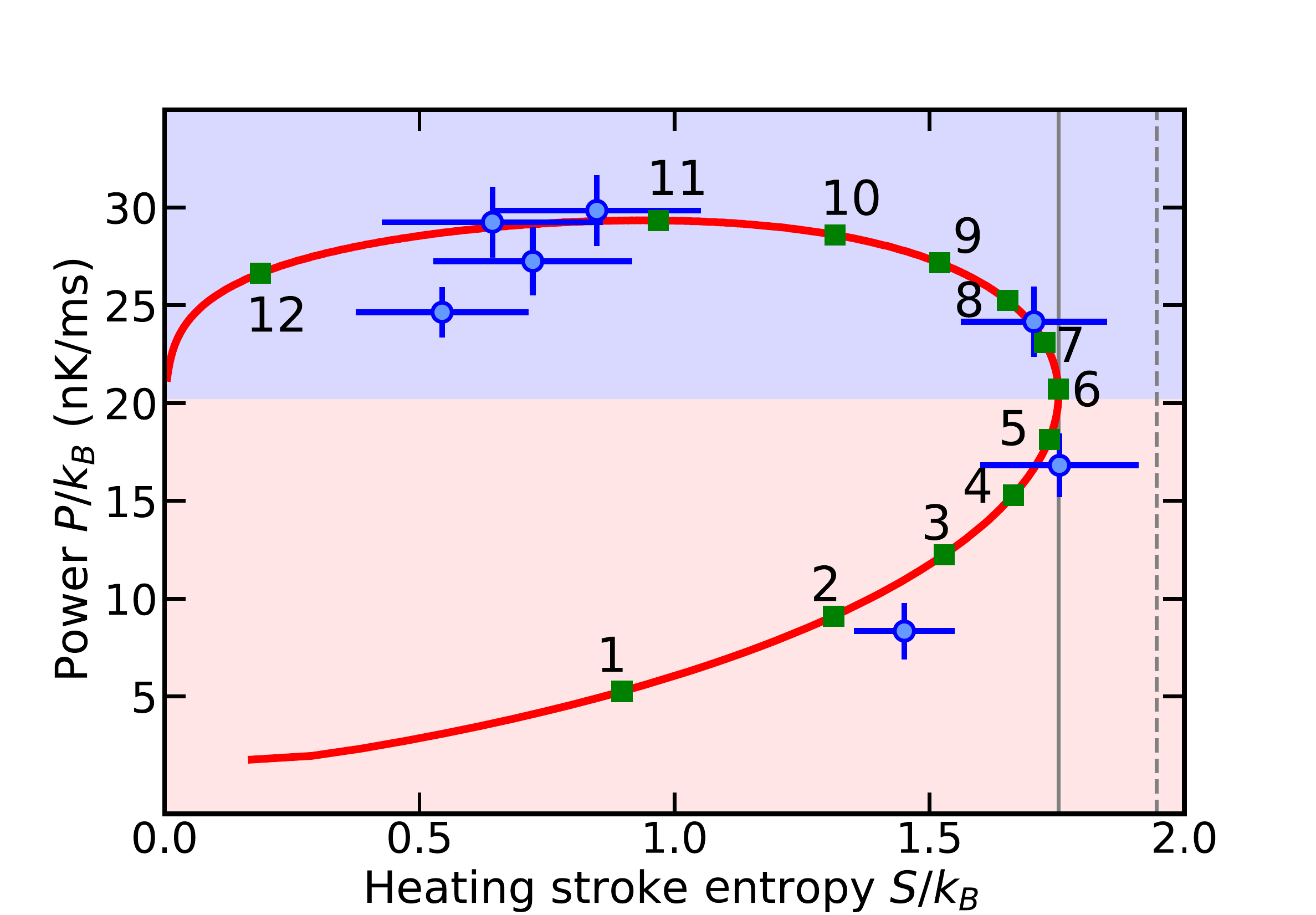}	
	\caption{Evolution of the power output. The blue dots represent the experimental data and the red line is the theoretical prediction based on our spin-exchange model. Green dots and associated numbers give the number of total spin-exchange collisions for a full cycle (Collision number~twelve is not located at the end of the red line due to numerical errors in calculating the total number of spin-exchange collisions of approximately $1\%$). The red area marks the positive temperature regime and blue stands for negative temperatures, when most atoms are populated in energetically higher states. 
	{The grey dashed vertical line indicates the system's theoretical entropy maximum $\ln(7)$ and the solid vertical line the engine's maximum entropy at $1.75$. Maximum power output is reached at roughly half that value.}}
	\label{fig:PowerFluctuations}
\end{figure}

\begin{figure}[t]
	\centering
	\includegraphics[width=8.2cm]{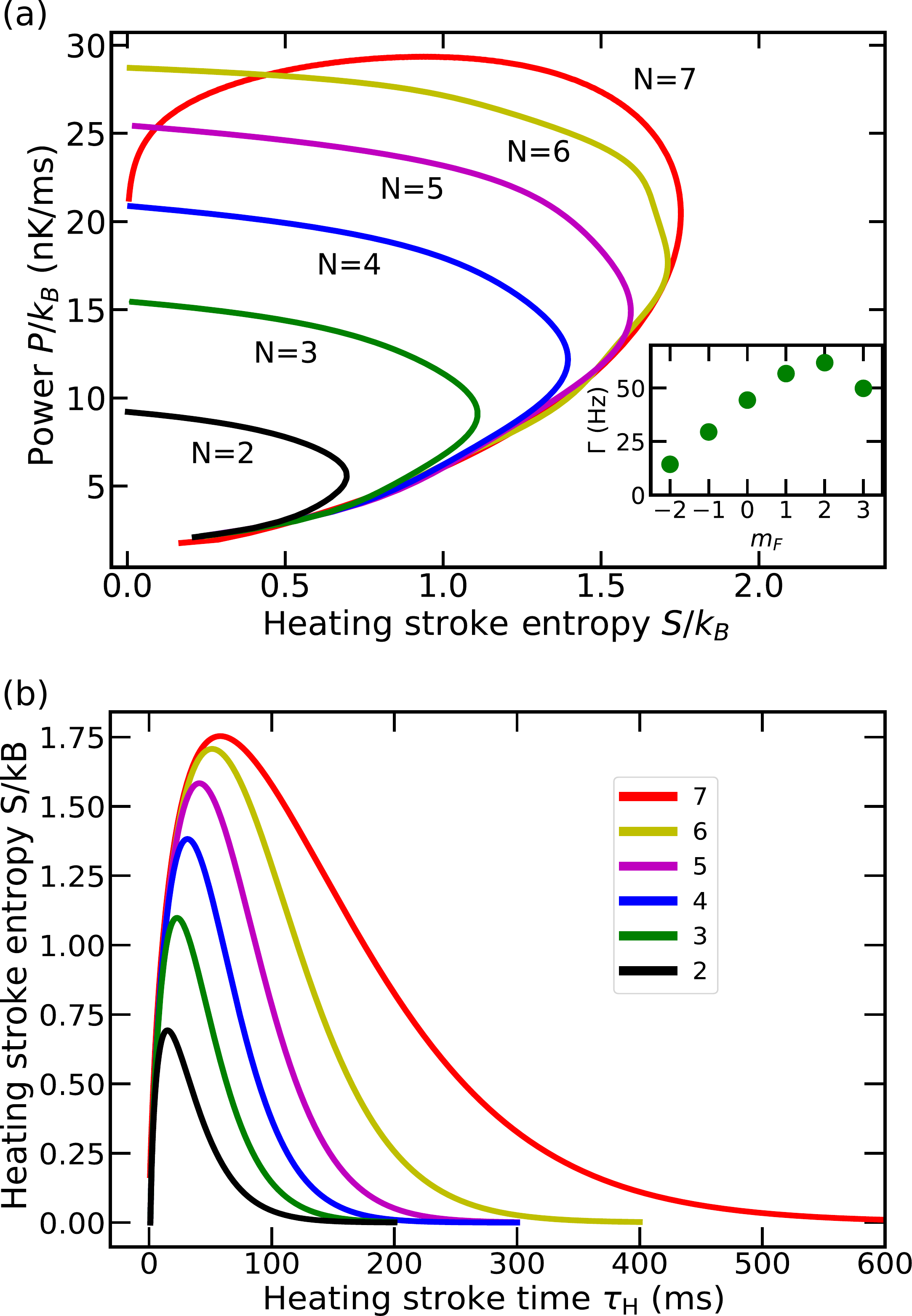}	
	\caption{(a) Evolution of the power-output  for truncated working medium with $N\leq 7$ levels as a function of the heating stroke Shannon entropy. Heat exchange \el{rate} and cycle time are considered to be the same as for the $(N=7)$-level heat engine. Inset shows the rates between two neighboring levels during the heating stroke. The $x$-axis gives the original state, i.e., $\Gamma^{3 \rightarrow 2}$, located at $m_{F,\text{Cs}}=3$, up to $\Gamma^{-2 \rightarrow -3}$, located at $m_{F,\text{Cs}}=-2$. (b) Evolution of the corresponding Shannon entropy as a function of the heating time $\tau_\text{H}$.}
	\label{fig:PowerFluctuations}
\end{figure}

\section{Power enhancement}
We next analyze how negative spin temperatures  and quantum fluctuations affect the power output of the heat engine. The efficiency $\eta$,  defined by the ratio of work output and heat input, is always constant since driving is  adiabatic in our experiment \cite{bou20}. As a result, nonadiabatic transitions do not occur during expansion and compression phases. The power output $P$ is given by the work output $|\langle W\rangle|= \langle Q_H\rangle -|\langle Q_C\rangle |$ divided by the cycle time
\begin{align}
P= \frac{\langle Q_H\rangle-\langle|Q_C|\rangle }{\tau_\mathrm{cycle}}=\frac{|\langle W\rangle|}{\tau_\text{cycle}}.
\end{align}
We follow the evolution of the power output during the heat engine cycle by plotting $P$ as a function of the Shannon entropy $S$ {during the heating stroke}  as shown in Fig.~\ref{fig:PowerFluctuations}.  It is   sufficient to consider only the engine's heating stroke entropy, since the final heating stroke entropy determines the initial cooling stroke entropy (the compression step being unitary), and the Cs ground $\ket{m_{F,\text{Cs}}=3}$ has to be reached on the cooling stroke side to realize a cycle. Blue dots represent experimental data and the red line is the theoretical simulation. The numerated green dots indicate the number of  spin-exchange collisions (associated with a transition from one discrete energy level to another).  The engine starts from the lowest energy state, where both {heating stroke} entropy and power are low. 
For increasing energy intake during the spin-exchange process, both  entropy and power increase very similarly  to the behavior  of standard thermal engine. The spin temperature is positive (red shaded area) until the point of maximum entropy (i.e.~maximum fluctuations)  {$S_\mathrm{max}/k_\text{B} = 1.75 \approx \ln 7$}   is reached after roughly 6 spin-exchange collisions. Subsequently, the engine operates  in the effective negative-temperature regime (blue shaded area). Here the Shannon entropy saturates and then starts decreasing,  while the power output further increases by roughly $30 \%$ compared to the case at maximum heat-stroke entropy. 
The maximum power of $P_\mathrm{max}=k_\text{B}\times 30\,$nK/ms is reached at {roughly} half the maximum entropy of $S/k_\text{B} \approx 0.9$,
corresponding to eleven spin-exchange collisions during the cycle. The power decreases after this point, since the work output saturates and the cycle time keeps increasing. These results clearly demonstrate the power boost associated with  negative spin temperatures.\\

\section{Comparison with a $N$-level engine}
We next numerically compare the thermodynamic performance of our quasi-spin  $(N=7)$-level quantum heat engine to that of a $(N\leq 7)$-level machine in order to analyze the influence of the dimension of the Hilbert space on the power output and on the size of the quantum fluctuations in the system \cite{den21}. To that end, we simulate the behavior of the $N$-level system by truncating the system of rate equations which governs the evolution of the populations of the complete $N=7$ system to a variable number $N\leq7$, by keeping the transition rates unchanged (see Appendix A, in particular Eq.~\eqref{eq:differential_eq}). For each system, $m_{F,\text{Cs}}=3$ keeps representing the ground state, and the level spacing and environmental parameters are unchanged. This leads to systems with the same transition rates between neighboring states while the number of levels is reduced. Afterward, the corresponding heat exchange $Q_H=\sum_n E_n^\text{Cs} \Delta p_n$ is determined from the spin dynamics. To guarantee the same heating velocity, the corresponding cycle time of the seven-level system is used for the same amount of exchanged heat. 

Figure~\ref{fig:PowerFluctuations}({a}) shows the evolution of the power output (3) as a function of the heating stroke entropy for various numbers $N$ of levels. We note that both the value of the maximum entropy and of the maximum power output increase with the number $N$ of levels. A $(N\geq 2)$-level quantum heat engine thus outperforms a two-level machine (while displaying,  at the same time, larger quantum fluctuations). Figure~\ref{fig:PowerFluctuations}({b}) exhibits the evolution of the corresponding Shannon entropy (2). All $N$-level systems start with low power and entropy values.  Both quantities increase until an entropy peak.
\el{After peaking, higher energy states are populated until the population is completely localized in the highest energy state, resulting in increasing engine-spin polarization and thus entropy decrease to ideally zero. At the same time, the power is steadily increasing. }

Interestingly, we observe the same behavior for all reduced-level systems.
\el{The seven-level system overall shows the same behavior, but for the last part of the evolution, the power drops slightly as the entropy further decreases. We attribute this small power drop to larger cycle times due to the low rate $\Gamma^{-2 \rightarrow -3}$ that populates the $m_{F,\text{Cs}}=-3$ state, as shown in the inset.}

\section{Conclusion}
We have experimentally investigated the performance of a quasi-spin quantum Otto engine coupled to a quantum Rb bath in the regime of effective negative temperature. Employing  time-resolved population measurements of the quasi-spin Cs state, we have concretely examined the influence of the bounded discrete energy spectrum on both the effective spin temperature and on the quantum fluctuations in the engine (quantified with the Shannon entropy), during the entire thermodynamic cycle. We have observed the transition to negative effective temperatures when the Shannon entropy of the quasi-spin distribution reaches its maximal value (after six spin-exchange collisions), corresponding to maximum quantum fluctuations. We have additionally  found that the power output is maximal at half the maximum entropy (after eleven spin-exchange collisions). Overall, negative spin temperatures lead to a significant increase of the power output of the engine, up to 30$\%$ compared to positive spin temperatures for the parameters of our experiment. \el{Additionally, the natural interaction stop, with its polarizing effect in the outmost states, halves the entropy with respect to its peak value while the engine provides maximum power output.}
Our findings offer insight into the connection between the dynamics of quantum spin populations of a  heat engine and its thermodynamic properties at the level of individual quanta.

Acknowledgements. We thank E. Tiemann for the provided scattering cross-sections underlying our spin-exchange model.
This work was funded by Deutsche Forschungsgemeinschaft (DFG) via Sonderforschungsbereich (SFB) SFB/TRR 185 (Project No. 277625399) and Forschergruppe FOR 2724, SB acknowledges funding by the Studienstiftung
des deutschen Volkes.
\bibliography{bibliography}

\appendix*

\section*{Appendix A: Temperature fit}
\label{sec:Temperature_fit}
\begin{figure}[h!]
	\centering
	\includegraphics[width=9cm]{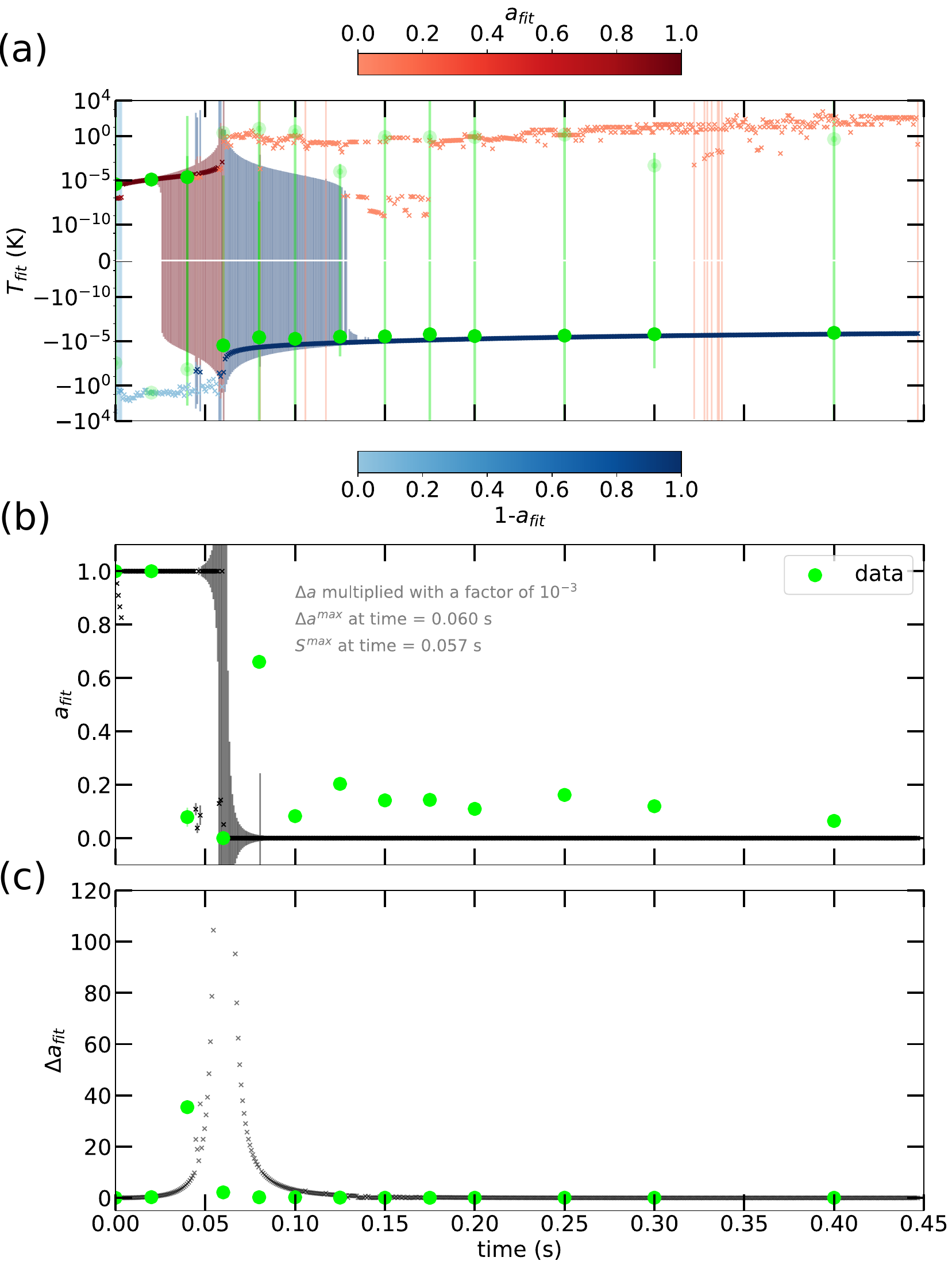}
	\caption{{Cs population fitted by a combination of Boltzmann distributions with negative and positive temperatures.}  
	(a) Fitted temperature over heating time. Our engine shows a thermal behavior for short times until it reaches the transition from positive to negative temperature at roughly $60\,$ms, corresponding to maximum entropy. Further away from this area, the system can be described by an almost exclusively positive (negative) temperature for short (long) heating times. (b) and (c) presents the fit parameter $a$ with its width $\Delta a_{fit}$. Green dots show data, colored lines simulations.}
	\label{fig:Combined_Plots_T_a}
\end{figure}

{During heating and cooling strokes, the Cs population can be inverted, that is, the majority is populating energetically lower or higher states. The heating stroke starts in the ground state $\ket{m_{F,\text{Cs}}=3}$ and is populating energetically higher states. To extract the temperature of the Cs spin population, we use a combination of two Boltzman distributions with position and negative temperatures, respectively. The fitted population writes $P_n = a  P_{+} + (1-a) P_{-}$ where \el{$P_\pm = \exp(-E_n^\text{Cs}/k_B T_\pm)/Z$}, with the partition function Z, as well as positive ($T_+$) and negative ($T_-$) temperatures. The fit parameter $a$ gives the contribution of each temperature regime. Figure \ref{fig:Combined_Plots_T_a} (a) shows the two temperature regimes for different heating times with its transition to negative temperatures. The color code represents the weighted influence of each temperature regime. This transition can also be seen by looking at the behavior of (b) the fit parameter  $a$ and (b) of the corresponding width $\Delta a$. Green dots show data, colored lines simulations based on our spin-exchange model (see Appendix B).}

\section*{Appendix B: Spin-exchange model}
\label{sec:Temperature_fit}
\begin{figure}
	\centering
	\includegraphics[width=4.5cm]{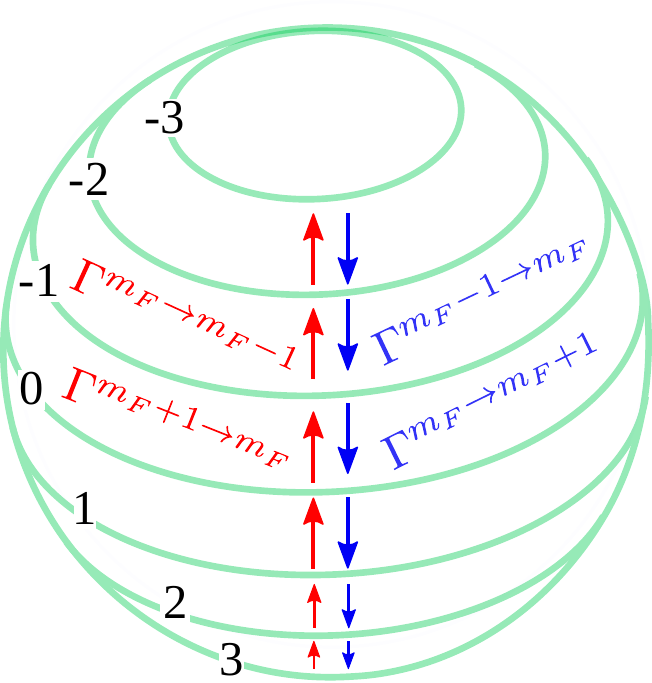}
	\caption{{{Cs spin-exchange rate illustration.} The Bloch sphere illustrates the seven-level Cs spin system with corresponding $m_{F,\text{Cs}}$ numbers. Red (blue) arrows present rates between two neighboring states for the heating (cooling) stroke.}}
	\label{fig:Rate_equation_illustration}
\end{figure}

To model the seven-level Cs system, we simulate the spin dynamics using a rate equation, given in Eq.~\eqref{eq:differential_eq}. The vector ${P}_{m_{F}}$ gives the population. The transition rates between two neighboring states are presented as a matrix. Each rate is calculable as \cite{bou20}
\begin{align}
{\Gamma^{m_F} = \braket{n} \, \sigma_{m_{F}} \, \bar{v}.}
\label{eq:Rates}
\end{align}
{Here $\sigma_{m_{F}}$ gives the cross-section, $\braket{n}= \int{n_{\mathrm{Cs}}(\vec{r}) \, n_{\mathrm{Rb}}(\vec{r}) \, d\vec{r}}$ the density overlap between the two atomic species and $\bar{v} = [(8 \, k_B \, T)/(\pi \, \mu)]^{1/2}$ the relative velocity with reduced mass $\mu$.}  
{Figure~\ref{fig:Rate_equation_illustration} shows the seven-level system with corresponding transition rates. For the cooling (heating) stroke, the bath polarizes the $\ket{m_{F,\text{Rb}}}=1$ ($\ket{m_{F,\text{Rb}}}=-1$) state. Processes that populate higher (lower) energetic Cs states are energetically forbidden, that is, red (blue) indicated rates are negligible.}

\newpage

\begin{widetext}
	\begin{equation}
	\label{eq:differential_eq}
	\begin{split}
	{\dot{P}_{m_{F}}  = \left( \begin{matrix} 0 & -\Gamma^{m_{F} \rightarrow m_{F}+1} & 0 & \\ \Gamma^{m_{F}+1 \rightarrow m_{F}} & 0 & \Gamma^{m_{F}-1 \rightarrow m_{F}} & \\ 0 & -\Gamma^{m_{F} \rightarrow m_{F}-1}  & 0 & \hdots\\ &\vdots&  & \end{matrix} \right) \cdot  
	\left( \begin{matrix} P_{m_{F}+1}\\P_{m_{F}}\\P_{m_{F}-1}\\ \vdots \end{matrix} \right)} .
	\end{split}
	\end{equation}
\end{widetext}

-----------

\end{document}